\documentclass[aps,unsortedaddress]{revtex4-2}
\usepackage[T1]{fontenc}
\usepackage[utf8]{inputenc}
\usepackage{xcolor,graphicx,bm,lineno}
\usepackage{tabularx}
\usepackage{booktabs}
\usepackage{float}
\usepackage{xcolor}
\usepackage{changepage}

\makeatletter

\renewcommand{\p@subsection}{}
\makeatother

\begin{document}

\title{Sensing, Traffic, and Construction in Termites}

\author{Yufei Xiao}
\thanks{These authors contributed equally to this work.}

\author{Qinglin Wu}
\thanks{These authors contributed equally to this work.}
\affiliation{School of Chemical and Biomolecular Engineering, Georgia Institute of Technology, USA\\
BioFrontiers Institute and Department of Chemical and Biological Engineering, University of Colorado Boulder, USA}

\author{Kukhyun Lim}
\thanks{These authors contributed equally to this work.}

\author{Nan-Yao Su}
\affiliation{Institute of Food and Agricultural Sciences, Department of Entomology and Nematology, Fort Lauderdale Research and Education Center, University of Florida, USA}

\author{Paul Bardunias}
\affiliation{Department of Biological Sciences, Florida Atlantic University, USA}

\author{Atanu Chatterjee}
\email{atanu.chatterjee@colorado.edu}

\author{Saad Bhamla}
\email{saad.bhamla@colorado.edu}
\affiliation{School of Chemical and Biomolecular Engineering, Georgia Institute of Technology, USA\\
BioFrontiers Institute and Department of Chemical and Biological Engineering, University of Colorado Boulder, USA}

\begin{abstract}
Subterranean and mound-building termites excavate, transport, and build within the same granular substrate that later regulates how they sense, move, and deposit material. From antennal-scale contacts through body-scale traffic to meter-scale architecture, this review synthesizes three linked problems: how workers sense local geometry and physical cues during search and excavation; how traffic moves through narrow, evolving conduits; and how excavation and deposition remodel the substrate that guides later behavior. Across these length scales, noisy local interactions couple sensing, transport, and construction through a shared material medium, leading to emergent order at the colony scale. We emphasize what is established experimentally, where evidence remains sparse or limited to a few model systems, and how emerging imaging, tracking, and modeling tools are making these feedbacks quantitatively accessible. We use this synthesis to motivate a quantitative physics-of-life framework for termite colonies that continually rewrite the medium through which they sense, move, and build.
\end{abstract}


\maketitle


\section{Introduction}
Much of active matter theory examines how energy-consuming agents move, interact, and organize in fixed geometries or under well-defined boundary conditions~\cite{marchettiHydrodynamicsSoftActive2013,bechingerActiveParticlesComplex2016,shaebani2020computational}. Termite collectives present a different mesoscale problem: sensing and decision-making, transport and traffic, and construction and repair all unfold within the same granular medium, and each leaves traces that shape what happens next. In termites, collective motion occurs in confinement that the colony continually rewrites~\cite{ziepke2022multi,dias2023environmental,nguyenWherePhysicsMeets2025}.

This is especially clear in subterranean termites. Most workers operate underground in complete darkness, in colonies that can reach up to a million individuals. They navigate through soil using touch, vibration, moisture, and chemical cues rather than vision~\cite{tranielloBehaviorEcologyForaging2000}. Food is patchy and, for buried resources, often cannot be detected reliably over useful distances through the substrate~\cite{pucheApplicationFractalAnalysis2001,pucheTunnelFormationReticulitermes2001}. Colonies search by excavating, and the same workers that extend the tunnel network also transport soil, maintain passages, regulate exposure and moisture, and sustain bidirectional traffic as local conditions change~\cite{korbThermoregulationVentilationTermite2003,kingTermiteMoundsHarness2015,michaelFindingShortcutsCollective2023}. Soil is at once the terrain they move through, the material they build with, and part of the medium through which information travels.

Each act of excavation, deposition, or repair modifies the substrate encountered by later workers. Coordination appears to rely less on long-range diffusing signals~\cite{fouquetCoordinationConstructionBehavior2014,petersenArrestantPropertyRecently2015,greenExcavationAggregationOrganizing2017} than on persistent changes in geometry together with shorter-lived local fields, including crowding, vibration, and substrate state. These stigmergic~\cite{grasseReconstructionNidCoordinations1959,theraulazBriefHistoryStigmergy1999} cues couple ongoing activity to prior environmental modification without centralized control, making termite colonies a useful system for studying environmental memory in living active matter~\cite{bano2024controlled,katoOptimalityTheoryStigmergic2025}.

\begin{figure}[t]
    \centering
    \includegraphics[width=1\columnwidth]{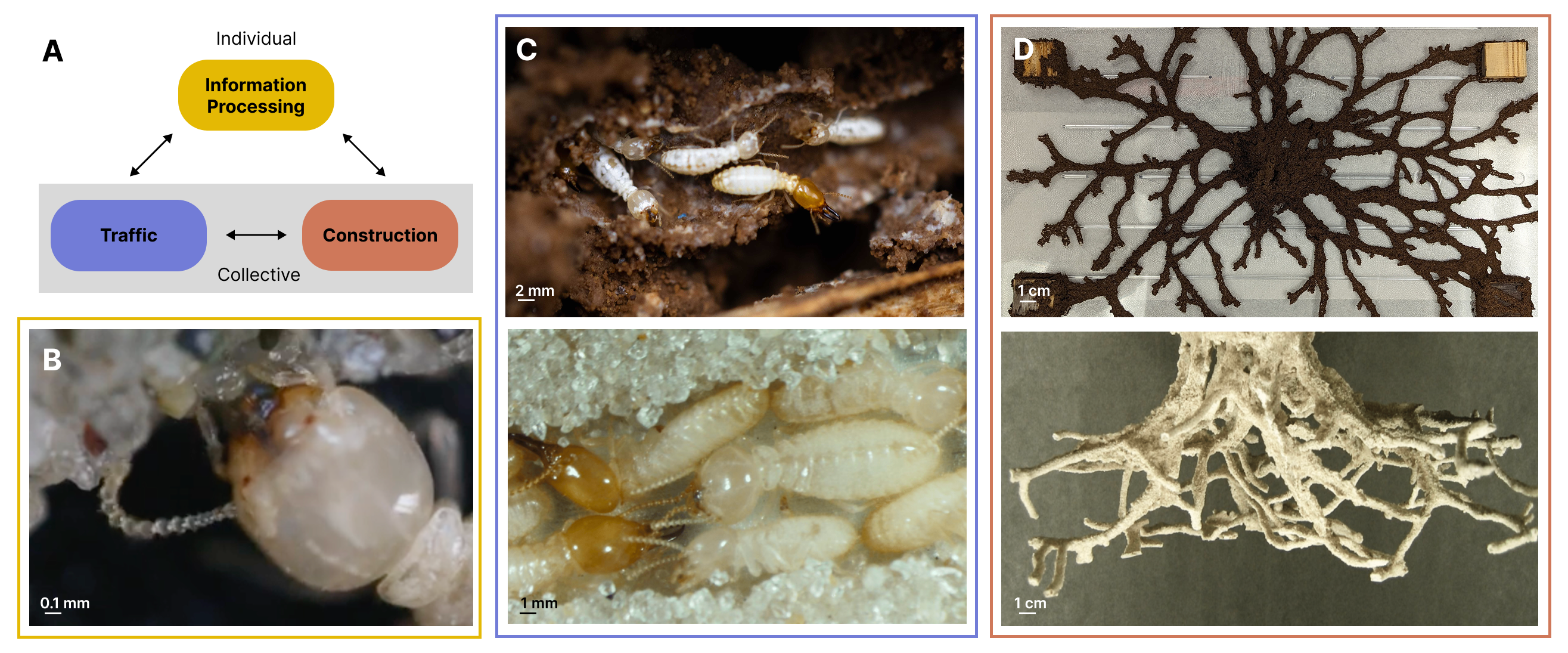} 
    \caption{\textbf{Overview of the physical and behavioral themes developed in this review.} (A) Schematic linking information, movement, and construction in subterranean termite colonies. (B) antennal sensing, highlighting antennation as a primary mode of local information acquisition and interaction (photo: Qinglin Wu). (C) bidirectional traffic flow through confined conduits, shown in an above-ground shelter tube (top) and a below-ground planar tunnel network (bottom) (photo: Qinglin Wu). (D) construction across scales, from above-ground shelter tubes (top, photo: Kukhyun Lim) to extended below-ground tunnel networks (bottom; photo: Paul Bardunias).}
    \label{fig:figintro}
\end{figure}

In this review, we organize the discussion around three closely linked problems: how workers sense the substrate during search and excavation, how traffic moves through narrow and changing conduits, and how excavation and deposition remodel the boundaries that guide later behavior (Fig.~\ref{fig:figintro}). We focus on mechanistic studies that connect local interaction rules to evolving geometry, traffic, and construction, drawing mainly on subterranean termites, especially \textit{Coptotermes formosanus} and \textit{Reticulitermes} spp., together with a smaller body of work on mound-building taxa, much of it based on simplified tunnel assays~\cite{pucheApplicationFractalAnalysis2001,pucheTunnelFormationReticulitermes2001,lee_surface_2008,barduniasQueueSizeDetermines2010,greenExcavationAggregationOrganizing2017,caloviSurfaceCurvatureGuides2019,kingTermiteMoundsHarness2015,michaelFindingShortcutsCollective2023}.

\section{Sensing the Substrate and Decision-making}
In termites, workers sense a substrate that has already been modified by collective activity. The key problem is how local measurements of geometry, material conditions, and physical cues are converted into decisions under confinement and uncertainty. Excavation, deposition, widening, and branching leave traces that later workers can encounter directly. Some of these traces persist: tunnel width, curvature, roughness, and branching pattern can remain long after the workers that produced them have moved on. Others are more transient, including local crowding, moisture, and vibration. Reading the substrate in termites means sensing both persistent geometry and shorter-lived physical cues, which together bias where workers move, excavate, and deposit next.

\begin{figure}
    \centering
    \includegraphics[width=1\linewidth]{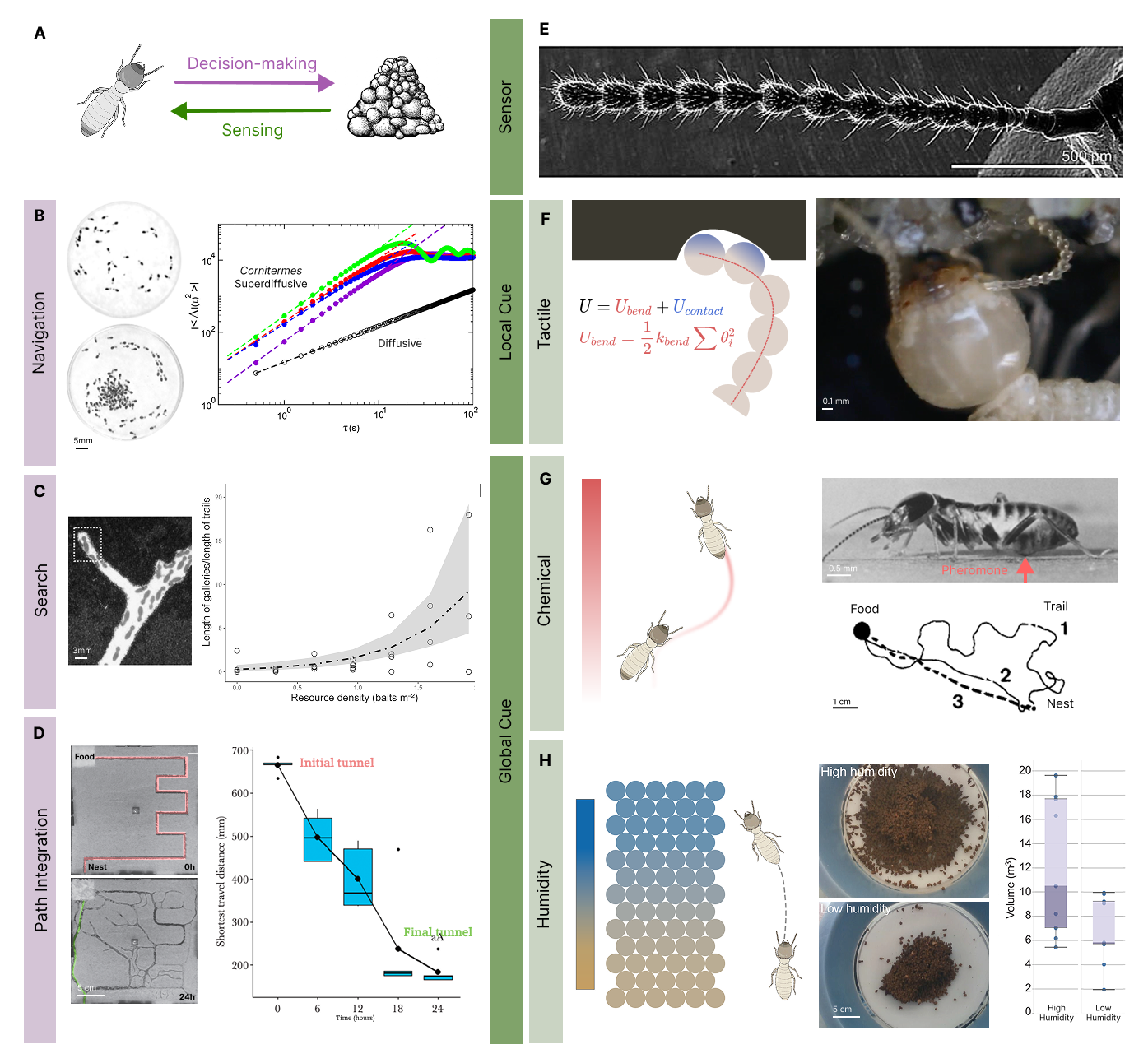} 
    \caption{\textbf{Information processing in termite colonies.} (A) Schematic of the coupled sensing-action loop between individual termites and the substrate they construct and inhabit. (B) Navigation. Mean squared displacement of four \textit{Cornitermes} sp. workers during open-arena foraging, with $\mathrm{MSD}\propto D \tau^\alpha$ indicating MSD scaling~\cite{miramontesLevyFlightsSelfSimilar2014}. For example, purple indicates $\alpha=1.90$. (C) Search. Left: Close-up of termite collective excavating tunnels~\cite{leeSimulationStudyTunnel2010}. Right: tunnel-tip activity and correlation between resource density and total tunnel length in \textit{C.~formosanus}~\cite{Almeida_Cristaldo_Desouza_Bacci_Florencio_Cruz_Santos_Santana_Oliveira_Lima_etal._2018}. (D) Path optimization. Left: tunnel network of \textit{C.~formosanus} shows pruning of long paths and reinforcement of short paths from the nest (bottom) toward a food source (top). Right: Shortest path distance decrease over time~\cite{michaelFindingShortcutsCollective2023}. (E) Moniliform antenna of \textit{Coptotermes formosanus}~\cite{yanagawa2009classification}. (F) Tactile sensing of local geometry. Left: schematic of antenna deforming against substrate curvature at the tunnel tip. Right: mechanosensillum and worker probing the tunnel face during excavation (Photo: Qinglin Wu). (G) Effect of chemical field. Left: Schematic of pheromone trail following. Right: (Top) \textit{Hodotermes mossambicus} worker laying a pheromone marking, (Bottom) Successive foraging paths of a single worker, with progressive path shortening consistent with pheromone trail reinforcement~\cite{heideckerOrganisationCollectiveForaging1984}. (H) Effect of moisture field. Left: schematic of moisture-mediated sensing. Right: mound morphology of \textit{Macrotermes michaelseni} in high- (top) and low-moisture (bottom) settings~\cite{Carey_Calovi_Bardunias_Turner_Nagpal_Werfel_2019}.}
    \label{fig:figinfo}
\end{figure}

\subsection{Sensory channels in confined environments}
For termites, sensing is a problem of information acquisition under confinement. Workers probe a narrow, deformable, and self-constructed medium whose geometry both constrains motion and carries information about prior activity (Fig.~\ref{fig:figinfo}A). The relevant signals are predominantly physical: wall contact, substrate-borne vibration, local moisture, and contact with nestmates. Each modality is noisy and local, so useful decisions must be aggregated from partial measurements distributed across time, body posture, and movement~\cite{tkacikInformationProcessingLiving2016,Khodak_Sulger_ODonnell_2016,Howarth_Moldovan_2018}.

\paragraph*{Antennal sensing as geometric sampling.} In \textit{Coptotermes formosanus}, scanning electron microscopy reveals at least nine distinct sensilla types distributed along the moniliform antenna, with mechanosensory sensilla on most segments and chemosensory sensilla concentrated distally~\cite{castilloComparativeAntennalMorphometry2021}. Beyond this anatomical inventory, the antenna can be viewed physically as a slender, segmented tactile probe whose deformation depends on how it is driven into contact with the substrate. Like other active tactile organs, its signals are generated by both external structure and the organism's own motion. The star-nosed mole (\textit{Condylura cristata}), for example, reconstructs burrow geometry through rapid sequential touch~\cite{cataniaEarlyDevelopmentSomatosensory2001a}, while cockroaches use antennal deflection to distinguish wall contact from other perturbations~\cite{chapmanModelAntennalWallfollowing2006}. In each case, the informative variable is the time-dependent pattern of contact, bending, and self-motion.

The termite antennae's moniliform morphology differs from the geniculate antennae of ants and the filiform antennae of cockroaches and basal termite relatives~\cite{chapmanModelAntennalWallfollowing2006}. Functionally, this beaded chain morphology suggests two complementary modes of geometric sampling (Fig.~\ref{fig:figinfo}E). Pressed into a surface, the antenna conforms to local relief and can register features at the scale of grain size and divot depth. Swept across a surface, it senses curvature at the scale of the tunnel cross-section. Workers use both modes by sweeping the tunnel face before excavation and probing the walls while moving within structures~\cite{lee_surface_2008,barduniasQueueSizeDetermines2010}. The antenna may mechanically transduce local geometry into a pattern of contacts that biases where to bite, where to deposit, and how to move through confined space. While direct measurements linking antennal kinematics or contact patterns to excavation decisions are still lacking, the antenna can be viewed as an actively driven filament interacting with a disordered boundary, where contact and geometry shape the sensed signal.

\paragraph*{Moisture, vibration, and chemical fields.} Additional channels complement tactile geometry sensing by providing information about disturbance, substrate state, and route stabilization. Moisture is behaviorally relevant because most termite workers have soft, unsclerotized cuticles that lose water rapidly in dry air~\cite{tranielloBehaviorEcologyForaging2000}. Humidity gradients can bias tunnel routing and trigger shelter-tube construction during above-ground foraging~\cite{suTunnelingActivitySubterranean2003}.

High-moisture sand also increases excavation output in \textit{C.~formosanus}~\cite{suTunnelingActivitySubterranean2003}, measured as the volume of sand removed over time, establishing that moisture modulates digging rate and not only path choice. Workers transport moist soil from excavation sites onto adjacent dry surfaces, physically separating deposition from excavation~\cite{Carey_Calovi_Bardunias_Turner_Nagpal_Werfel_2019}. Because deposited wet soil can create local regions of elevated relative humidity that persist after the deposition event, moisture may function stigmergically as a spatial template that marks active build zones rather than serving only as a physiological constraint~\cite{Carey_Calovi_Bardunias_Turner_Nagpal_Werfel_2019}. Cross-species field data support a related environmental effect: in Eastern Africa, \textit{Macrotermes michaelseni} and \textit{M.~natalensis} build larger mounds in higher-moisture ecosystems (Fig.~\ref{fig:figinfo}H), suggesting that ambient moisture scales construction output across taxa and environments~\cite{Carey_Calovi_Bardunias_Turner_Nagpal_Werfel_2019}. Airflow may provide a complementary orienting cue because its direction and speed define a persistent vector field that can guide growth without relying on a decaying chemical gradient.

Substrate-borne vibration carries a different class of information. Head-banging against tunnel walls propagates alarm signals through the soil, while longitudinal body oscillation appears to amplify local activity without specifying a particular action~\cite{stuartAlarmDefenseConstruction1967,sillam-dussesAlarmCommunicationPredates2023,sansomTermiteVibrationSensing2025}. Workers of \textit{Cryptotermes secundus} also detect vibrations produced by approaching ant predators~\cite{oberstCrypticTermitesAvoid2017}. Chemical cues, particularly trail pheromone from the sternal gland, can bias path choice at branch points and help stabilize established foraging routes (Fig.~\ref{fig:figinfo}G)~\cite{kaibTrailfollowingTermitesEvidence1982,reinhardSystematicSearchFood1997}. 

Together, these channels differ in range, persistence, and function. Within tunnels, where movement is already strongly constrained by geometry, tactile and geometric cues appear especially important for guiding excavation (Fig.~\ref{fig:figinfo}E). Chemical cues become more important when routes must be reinforced over larger distances, especially during above-ground foraging~\cite{reinhardSystematicSearchFood1997}.

\subsection{Search under uncertainty} 
Subterranean termite foraging is fundamentally a search problem under severe information constraints. Workers act on noisy, local measurements and typically have no direct access to the location of distant resources. Search proceeds through excavation itself: the colony generates information by probing and extending the substrate it inhabits (Fig.~\ref{fig:figinfo}C). Path integration, the accumulation of self-motion cues to maintain a position estimate relative to a reference point, is widespread among navigating animals, from desert ants to rodents~\cite{mullerPathIntegrationDesert1988}. In termites, tunneling \textit{C.~formosanus} workers maintain a heading away from the tunnel origin through idiothetic cues and correct their trajectory after forced lateral displacements~\cite{barduniasDeadReckoningTunnel2009, barduniasChangeTunnelHeading2010}. Passage through an open chamber resets this heading, indicating that the continuity of the tunnel itself contributes to the navigational signal~\cite{barduniasChangeTunnelHeading2010}. These findings imply that workers at the tunnel tip are, in effect, aware of the position of the tunnel origin. After food discovery, \textit{C.~formosanus} workers can construct shortcut tunnels connecting the tunnel origin directly to food~\cite{michaelFindingShortcutsCollective2023}, suggesting that workers at various points within the gallery system may be capable of geo-referencing their positions relative to both nest and food, although the underlying mechanisms remain unresolved.

Sensing provides a noisy signal~\cite{tkacikInformationProcessingLiving2016}, and workers act without direct knowledge of distant resource locations. For subterranean termites, this constraint is especially severe because buried wood is difficult to detect through soil at useful distances. In arena experiments with \textit{Reticulitermes flavipes}, nearby wood does not alter tunnel geometry until workers make physical contact, because soil attenuates volatile cues over distances as short as a few millimeters~\cite{pucheApplicationFractalAnalysis2001,pucheTunnelFormationReticulitermes2001}. Excavation is an active search process that generates information by probing the environment. In that sense, termite tunneling belongs to a broader class of search problems in which uncertainty is reduced through exploration itself~\cite{vergassolaInfotaxisStrategySearching2007}. The resulting movement statistics reflect this exploratory character. In open-arena foraging by \textit{Cornitermes} sp., mean squared displacement scales as $\mathrm{MSD}(\tau)\propto\tau^{\alpha}$, with $\alpha>1$ (Fig.~\ref{fig:figinfo}B), indicating superdiffusive movement in which localized exploration is interspersed with longer relocations~\cite{miramontesLevyFlightsSelfSimilar2014}.

\subsection{From local cues to excavation decisions}
At the tunnel tip, locally sensed cues are converted into decisions that directly modify the substrate. In \textit{C.~formosanus}, protrusions and concavities on the tunnel face trigger excavation or deposition depending on the scale of the feature: protrusions are preferentially removed and pellets are placed into concavities~\cite{lee_surface_2008}.

In \textit{Macrotermes} mound construction, convex surfaces likewise increase the likelihood of material removal, whereas concave surfaces increase the likelihood of deposition~\cite{caloviSurfaceCurvatureGuides2019}. This response can be framed quantitatively as an action probability or event rate that depends on local curvature at the scale of the tunnel tip or mound surface. Each construction event changes local curvature, and the modified curvature biases subsequent events, making construction a curvature-sensitive feedback loop written into the substrate (Fig.~\ref{fig:figinfo}E and~\ref{fig:figinfo}F).

Crowding behind an excavator provides a second input that has been shown to shift workers from waiting to lateral digging, contributing to tunnel widening and branch initiation~\cite{barduniasQueueSizeDetermines2010}. In \textit{C.~formosanus}, this response varies with queue size: queues of about seven workers excavated tunnels roughly twice as wide as queues of five or fewer workers, while excavation activity remained approximately balanced between the tunnel tip and lateral walls, at about $17$ bite events $\mathrm{min}^{-1}$ - approximately $500$ bite events over a $30$ min window~\cite{barduniasQueueSizeDetermines2010}. A related congestion-dependent reduction in participation has been observed in ants, where about $22\%$ of workers never enter the tunnel during collective excavation~\cite{aguilarCollectiveClogControl2018}. Together, these examples suggest that inactivity or lateral redistribution of work under confinement may be a shared response to narrow-tunnel congestion across social insects.

Because these same substrate features also constrain movement and encounters under confinement, the factors that guide excavation are closely linked to those that govern traffic. Geometry, roughness, and local density influence both action selection and where delays, queues, and passing conflicts arise. At the scale of the growing network, accumulated decisions self-organize into branching patterns that reduce overlap of explored territory, with tunnel orientations radiating outward from the nest in approximately uniform angular distributions~\cite{robsonNonrandomSearchGeometry1995,camporaTunnelOrientationSearch2001}. Effective spatial coverage thus emerges from locally coupled excavation and movement, without requiring long-range detection of resources.

\section{Traffic and Transport in self-constructed environments}
Termite colonies depend on continuous bidirectional movement within narrow structures such as tunnels and shelter tubes to deliver resources, export excavated soil, and coordinate work at distributed nests~\cite{barduniasOpposingHeadingsExcavating2009,barduniasQueueSizeDetermines2010}. Throughput depends on measurable features of the built environment, including tunnel width, curvature, roughness, body-to-tunnel size ratio, and encounter rate~\cite{sim2013measurement,sim_measurement_2012,lee_rounding_2008,ku_movement_2013}. Contacts and physical constraints generate queues, delays, and coordinated rearrangements that shape individual decisions and scale up into colony-level traffic patterns. Efficient transport and search within self-built structures therefore depend on resolving local conflicts without sacrificing collective throughput. However, most quantitative measurements to date come from controlled planar or otherwise simplified tunnel systems~\cite{lee_surface_2008,barduniasQueueSizeDetermines2010,mizumoto2020complex}, and models that extend these relationships to fully three-dimensional soil networks, where contact-dominated motion, geometry-sensitive transport, and substrate remodeling are coupled, remain sparse. 

\subsection{Why termite traffic is a distinct physics problem}
Traffic in termites differs from pedestrian systems and many ant trails because transport capacity is set primarily by short-range interactions under strong geometric confinement. Pedestrian flow and ant trails provide useful baselines for thinking about density, congestion, and counter-flow, but termite traffic occurs in narrow conduits whose geometry both constrains motion and is modified by use. Termite traffic thus occupies a regime in which transport and confinement coevolve. This transport problem can be described at several levels, from encounter-based models of individual motion to continuum and network flow. Box~1 summarizes that hierarchy and shows where termite traffic departs from standard fixed-boundary treatments.

\begin{figure}
    \centering
    \includegraphics[width=0.9\linewidth]{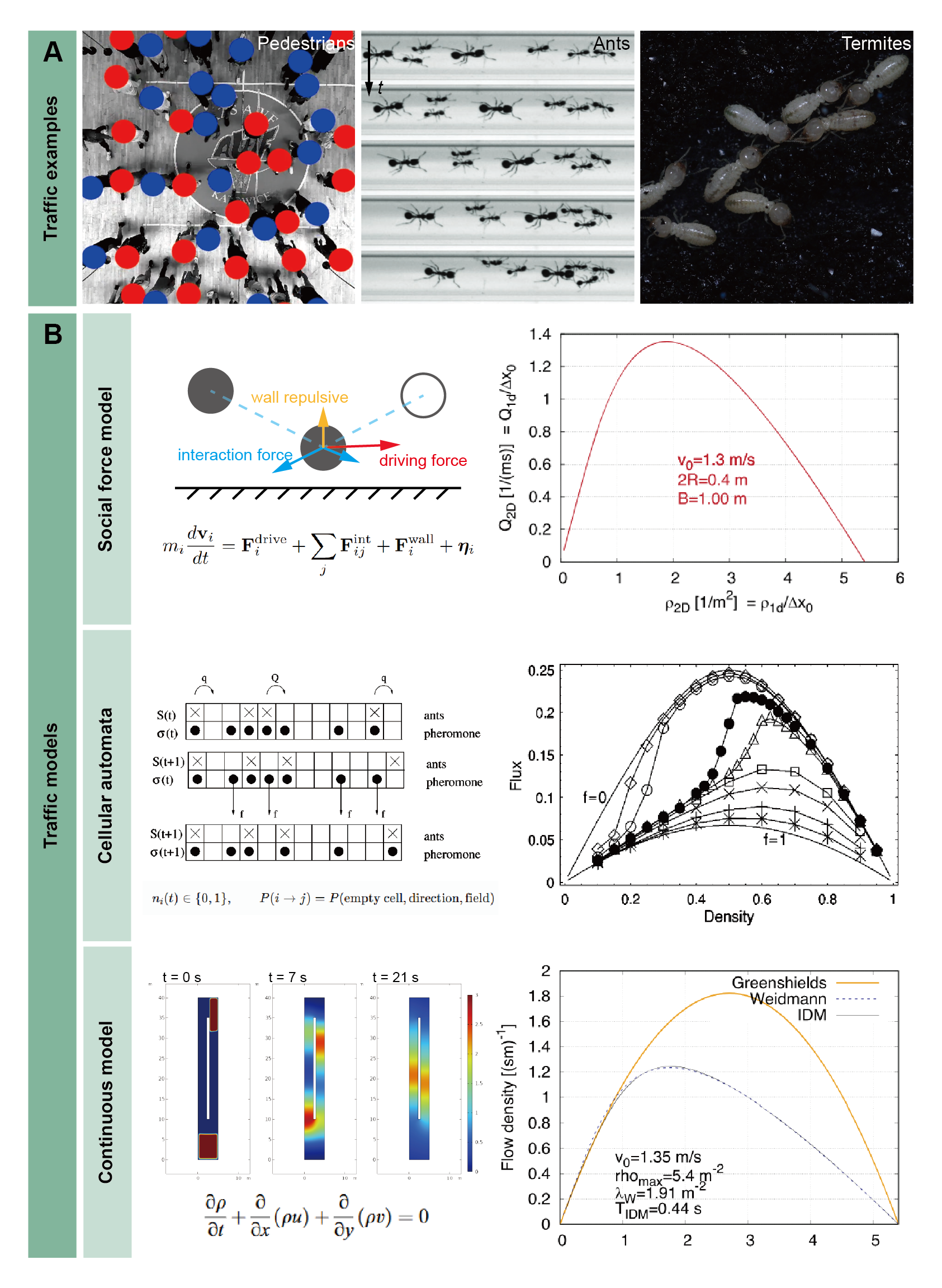}
    \caption{\textbf{Traffic systems and model classes.} (A) Representative traffic systems used as baselines for termite traffic: pedestrian crowds~\cite{bacik_lane_2023}, ant trails~\cite{gravish_glass-like_2015}, and termite movement in confined galleries (Photo: Qinglin Wu). (B) Modeling descriptions arranged by scale. Social-force models represent individuals as self-driven particles with driving, interaction, wall, and noise terms. Cellular automaton models represent occupancy and probabilistic movement on a lattice. Continuum models describe density and velocity fields through conservation laws. The corresponding flux-density relations summarize how transport capacity changes with density~\cite{helbing1995social,chowdhury_cellular-automata_2002,nishinari_cluster_2003,schmid_faster-than-real-time_2025,treiber_traffic_2025}.}
    \label{trafficmodel}
\end{figure}

\paragraph*{Interaction in confined geometry.} Termite movement is typically restricted to covered conduits with limited lateral degrees of freedom, so counter-flow conflicts are resolved mainly through tactile negotiation (wall-following, yielding, or turning) rather than through the long-range anticipation that often structures pedestrian flow~\cite{sim_measurement_2012,barduniasQueueSizeDetermines2010}. Transport resistance is often dominated by discrete encounter events and localized slowdowns, making traversal time, interaction time, and passing time especially informative observables~\cite{sim2013measurement,sim_measurement_2012}. Confinement also makes termites unusually sensitive to geometry. Because the boundary remains effectively fixed on short timescales, curvature, corners, constrictions, and surface irregularities can dominate travel time even when overall tunnel length remains unchanged~\cite{lee_rounding_2008,sim2013measurement,sim_measurement_2012,lee_effects_2023,ku_movement_2013}. Unlike many ant trails, whose effective width can expand through lateral spreading or lane formation, termite tunnels remain tightly constrained by boundaries the colony has already built.

\paragraph*{Movement in an adaptive structure.} The conduit itself can be modified during use. Congestion at excavation fronts can generate queues that bias lateral digging and hence tunnel widening~\cite{barduniasQueueSizeDetermines2010,barduniasOpposingHeadingsExcavating2009}. The balance between excavation and deposition can also influence branching and network topology~\cite{barduniasOpposingHeadingsExcavating2009}. Body mechanics contributes to traffic resolution as well, since the soft-bodied form of termites permits tight turning, reorientation, and squeezing in confined passages. These behaviors may shorten passing duration and resolve congestion differently from harder-bodied insects such as fire ants, in which U-turn strategies appear less common~\cite{gravish_glass-like_2015}. Termite traffic is flow through an adaptive conduit whose transport properties depend on the colony's own activity.

    \vspace{0.25in}
    \centerline{\textbf{Box 1\ \textbar\ Traffic theory across scales: from interacting walkers to adaptive conduits}}
    \par\noindent\rule{\linewidth}{0.4pt}
    
    \begin{adjustwidth}{10pt}{10pt}
    \small

    Traffic has been studied across many living and nonliving systems, from pedestrian crowds and vehicular lanes to ant trails, confined insects, and active particles under boundary constraints~\cite{helbing1995social,bellomo2011modeling,bechingerActiveParticlesComplex2016,couzin_self-organized_2003,dussutour_optimal_2004,chowdhury_cellular-automata_2002,nishinari_cluster_2003,treiber_traffic_2025}. These approaches differ mainly in what they take as the fundamental degrees of freedom: individuals, lattice occupancies, velocity distributions, densities, or network flows. Fig.~\ref{trafficmodel} provides a visual roadmap for this hierarchy, comparing pedestrian, ant, and termite traffic in panel A and organizing common model classes by scale in panel B. The schematic in Fig.~\ref{fig:trafficschematic} defines generic microscopic and macroscopic observables that can be extracted from confined bidirectional traffic.
    
        \begin{figure}[hb!]
        \centering
        \includegraphics[width=0.8\linewidth]{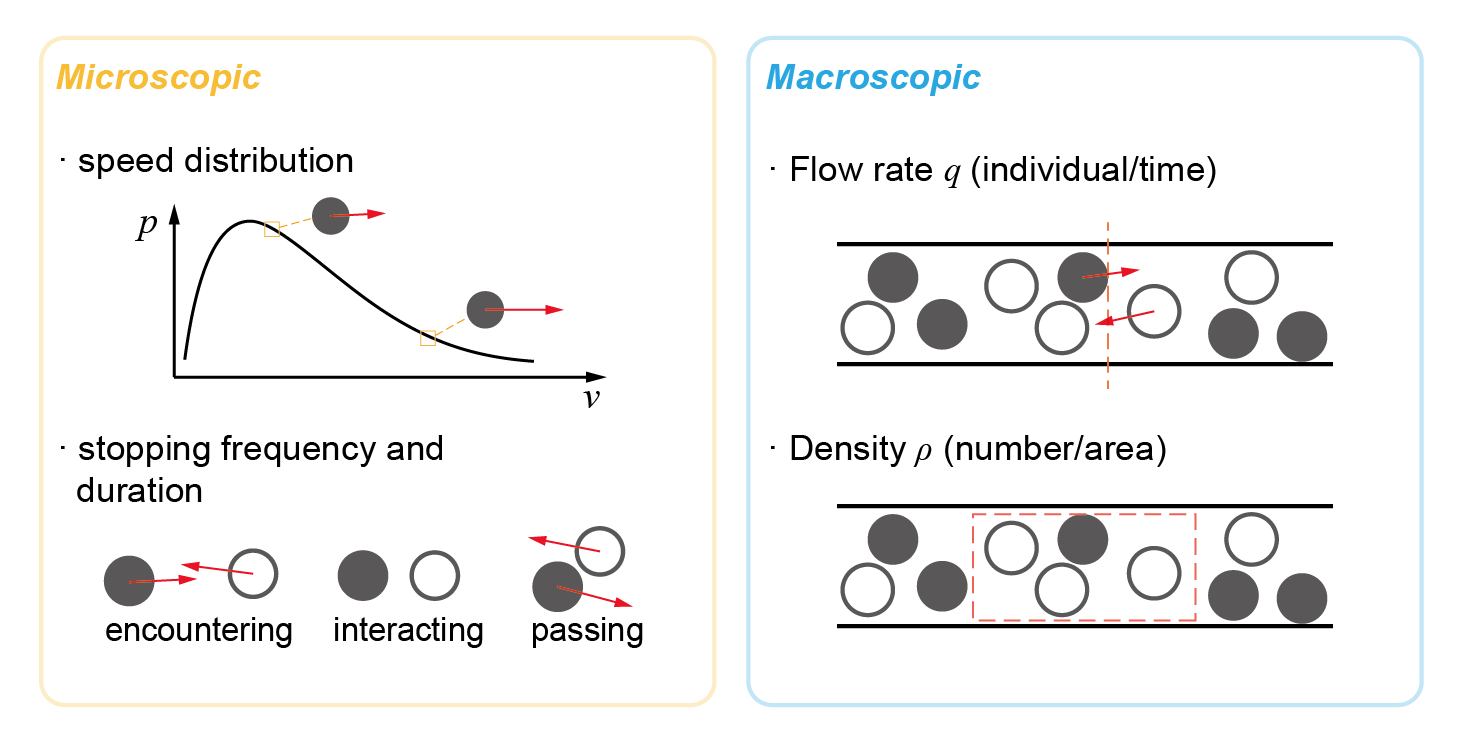}
        \caption{\textbf{Microscopic and macroscopic observables for confined bidirectional traffic.} (A) Microscopic observables extracted from individual trajectories include the speed distribution \(p(v)\), stopping frequency and duration, and transitions among encounter, interaction, and passing states during local contact events. (B) Macroscopic observables obtained by coarse-graining trajectories include flow rate \(q\), measured as the number of agents crossing a reference line per unit time, and density \(\rho\), measured as the number of agents within a defined area or channel length. Filled and open circles indicate agents moving in opposite directions.}
        \label{fig:trafficschematic}
        \end{figure}

    At the microscopic level, exemplified by social-force descriptions in Fig.~\ref{trafficmodel}B, individuals are represented explicitly and motion emerges from local interactions with neighbors and boundaries~\cite{helbing1995social,helbing_simulating_2000}:
    \begin{equation}
    m_i \frac{d\mathbf{v}_i}{dt}
    =
    \mathbf{F}^{\mathrm{drive}}_i
    + \sum_j \mathbf{F}^{\mathrm{int}}_{ij}
    + \mathbf{F}^{\mathrm{wall}}_i
    + \boldsymbol{\eta}_i 
    \end{equation}
    Such models are useful when transport is dominated by discrete encounters, including stopping, yielding, turning, and passing. The corresponding microscopic observables are those shown in Fig.~\ref{fig:trafficschematic}A: the speed distribution \(p(v)\), the frequency and duration of stops, and the probabilities of transitioning among encounter, interaction, and passing states during pairwise or local multi-body contact events. 

    A complementary agent-level description is provided by cellular automaton models, also shown in Fig.~\ref{trafficmodel}B. These models discretize space into occupied or empty sites and update movement through local rules for exclusion, directionality, and environmental fields such as pheromone. They are especially useful for ant-trail traffic, whereas termite applications would require update rules dominated by short-range tactile contact with walls and nestmates, head-on encounters, and local tunnel geometry, rather than by long-range visual anticipation.

    At a mesoscopic level, one instead tracks an ensemble distribution \(f(\mathbf{x},\mathbf{v},t)\) over individual positions and velocities~\cite{bellomo2011modeling}:
    \begin{equation}
    \partial_t f + \mathbf{v}\cdot\nabla_{\mathbf{x}} f
    + \nabla_{\mathbf{v}}\cdot(\mathbf{F}f)
    = C[f]
    \end{equation}
    This description bridges local encounter rules and collective observables such as density, speed, and flux. In confined biological traffic, \(f(\mathbf{x},\mathbf{v},t)\) could in principle be estimated from trajectory data, allowing one to ask how channel width, curvature, roughness, or boundary contact shifts the local velocity distribution, interaction time, and effective collision operator \(C[f]\).

    At larger scales, continuum descriptions such as those summarized in Fig.~\ref{trafficmodel}B treat traffic with a conservation law~\cite{lighthill1955kinematic,richards1956shock,schmid_faster-than-real-time_2025},
    \begin{equation}
    \partial_t \rho + \nabla\cdot\mathbf{J}=0
    \end{equation}
    together with a constitutive relation such as
    \begin{equation}
    \mathbf{J}=\rho\,\mathbf{u}
    \qquad\text{and}\qquad
    J=\rho\,u(\rho)
    \end{equation}
    The corresponding macroscopic observables include the density \(\rho\) and flow rate \(q\) shown in Fig.~\ref{fig:trafficschematic}B; these can be related to mean velocity \(\mathbf{u}\) and flux \(\mathbf{J}\) through continuum descriptions. In branching systems, a complementary network description treats channel segments as edges carrying flow,
    \begin{equation}
    J_e \sim \frac{\Delta \phi_e}{R_e}
    \end{equation}
    where \(\Delta \phi_e\) is an effective driving difference along edge \(e\), and \(R_e\) is an effective transport resistance. Each segment can also be assigned measurable geometric and dynamical properties such as width, curvature, travel time, roughness, and use frequency.

    These frameworks place termite traffic in context (Fig.~\ref{trafficmodel}A). Pedestrian systems highlight anticipatory crowd flow, ant trails illustrate route choice and counter-flow on branching paths~\cite{couzin_self-organized_2003,dussutour_optimal_2004,fourcassie_ant_2010}, and fire ants provide a close biological analogue for confinement-dominated transport~\cite{gravish_glass-like_2015}. Yet termites occupy a harder regime: movement is strongly contact dominated, but it occurs in conduits whose width, curvature, roughness, and connectivity have been constructed by the colony itself.
    
    This points to a further class of models in which transport is coupled to evolving geometry:
    \begin{equation}
    \partial_t g = \mathcal{F}(\rho,J,\sigma,\ell,\ldots)
    \end{equation}
    where \(g\) denotes a geometric property of the channel, such as width, roughness, curvature, or connectivity; \(\sigma\) denotes local contact or mechanical stress; and \(\ell\) denotes a queue length or waiting-state measure. In termites, this coupling can be stated in more specific terms. Local crowding or reduced throughput at a bottleneck may predict where tunnel widening occurs~\cite{barduniasQueueSizeDetermines2010}, opposing excavation fronts can bias branch initiation and network topology~\cite{barduniasOpposingHeadingsExcavating2009}, and geometric features such as curvature, constrictions, corners, or surface irregularities may increase encounter duration and passing delay over finite path lengths. The central termite-specific modeling problem is therefore traffic in a channel whose geometry is itself a dynamic variable shaped by repeated use.
    \end{adjustwidth}
    \par\noindent\rule{\linewidth}{0.4pt}

\begin{figure}
    \centering
    \includegraphics[width=1\linewidth]{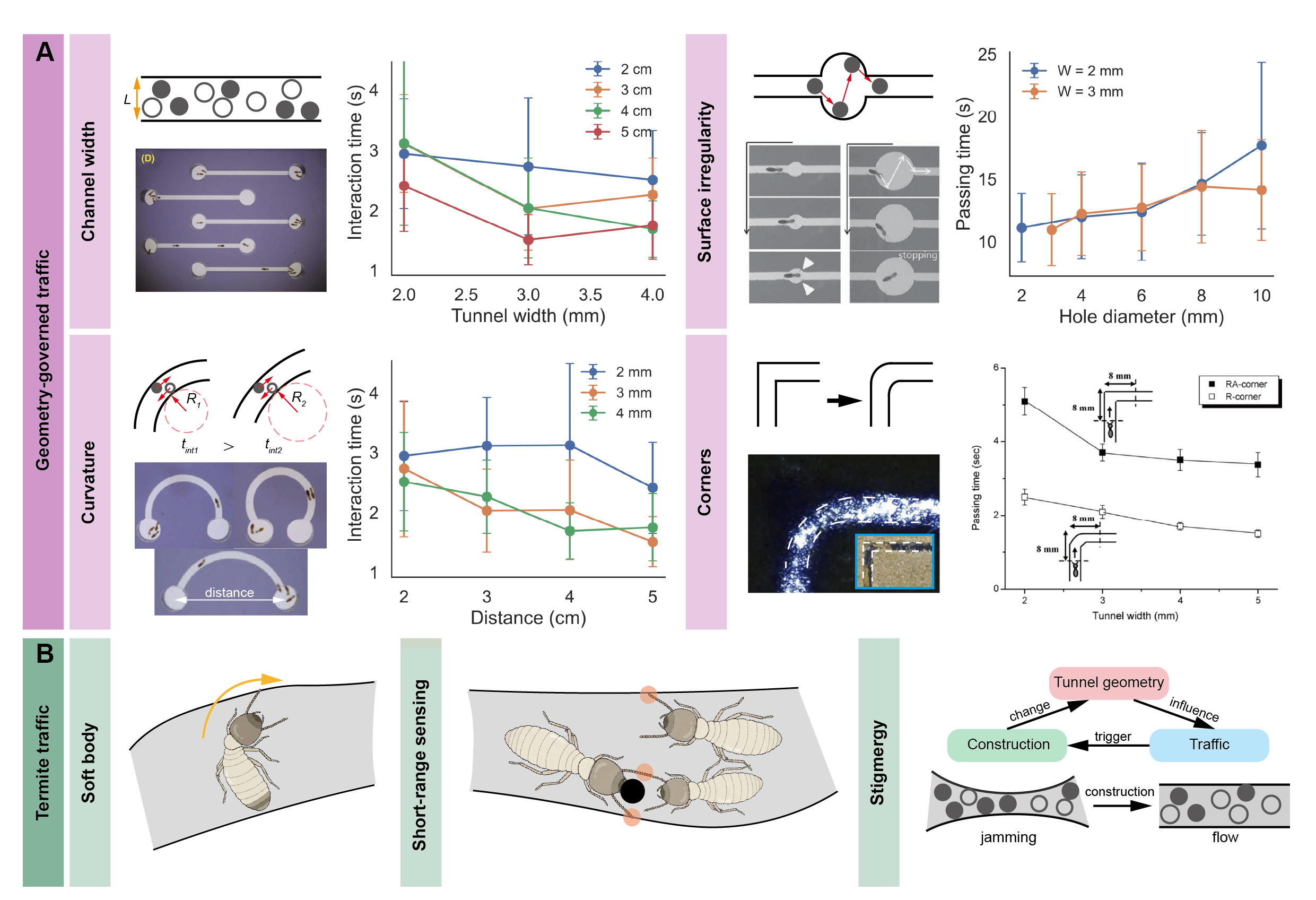}
    \caption{\textbf{Geometry-governed termite traffic.} (A) Controlled tunnel manipulations show that channel width, surface irregularity, curvature, and corners alter interaction or passing time. Wider channels generally reduce encounter delays, while irregularities, curvature, and sharp corners can increase transport costs; rounding or smoothing can reduce those costs~\cite{sim2013measurement,sim_measurement_2012,cho2014movement,lee_rounding_2008}. (B) Termite-specific features that distinguish traffic in self-built conduits: body compliance, short-range tactile sensing, and stigmergic remodeling, in which congestion can trigger construction that changes future flow.}
    \label{traffictermite}
\end{figure}

\subsection{Geometry, encounters, and transport costs}
While Fig.~\ref{trafficmodel} places termite traffic within broader traffic theory, Fig.~\ref{traffictermite} summarizes the termite-specific physical ingredients of the problem: geometry-dependent encounter costs, body compliance, short-range tactile sensing, and stigmergic remodeling of the conduit by traffic itself. In termite tunnels, transport cost is often set by discrete encounters shaped by width, curvature, corners, and surface irregularity~\cite{sim_measurement_2012,sim2013measurement,ku_movement_2013,lee_effects_2023}. These local interactions scale up into colony-level observables such as throughput, delay, and route use. Geometry defines the local transport landscape experienced by the colony.

\paragraph*{Width, bottlenecks, and self-organized widening.}  Tunnel width sets transport capacity because it determines whether termites can pass one another with minimal delay or must negotiate each encounter. Laboratory and field measurements indicate that foraging tunnel widths in \textit{C.~formosanus} commonly fall in the range of a few millimeters (approximately body width), though wider passages occur in some contexts~\cite{lee2020minimizing}. Direct behavioral measurements show that passing performance depends strongly on width and geometry when individuals encounter oncoming traffic (Fig.~\ref{traffictermite}A)~\cite{sim_measurement_2012,sim2013measurement}. At excavation fronts, queue formation can bias lateral digging and promote tunnel widening, linking congestion directly to capacity~\cite{barduniasQueueSizeDetermines2010}. Width is a variable that colonies adjust through work. Complementing fixed-width experiments, experiments with varied width along the tunnel path showed that spatial nonuniformity decreases movement efficiency~\cite{cho2014movement}.

\paragraph*{Curvature and surface irregularity.}  Other geometric features shape movement by altering how termites turn, stop, and negotiate encounters. Bends impose turning costs and create localized slowdowns. Termites move more efficiently through rounded corners in bent tunnels than through right-angled ones~\cite{lee_rounding_2008}. Workers also repeatedly deposited particles at corners, progressively transforming sharp bends into smoother paths. Construction thus reduced subsequent transport cost without any global planning. Curvature effects extend beyond a single bend. Using acrylic tunnels with controlled curvature, Sim et al.\ measured traversal and interaction times in tunnels of equal length and showed that both vary systematically with curvature (Fig.~\ref{traffictermite}A)~\cite{sim_measurement_2012,sim2013measurement}. 

Surface irregularity introduces a related class of geometric constraints. In controlled indentation experiments, irregularity was essential for rapid tunnel initiation, whereas its absence delayed tunneling for hours~\cite{lee_surface_2008}. In established tunnels, workers respond differently to irregularities of different geometries, sometimes exploring them and sometimes excavating them (Fig.~\ref{traffictermite}A)~\cite{lee_surface_2008,leeSurfaceIrregularityInducedtunneling2008,ku_movement_2013}. Building on these observations, recent modeling work incorporates stopping time at irregularity sites, together with the density and distribution of irregularities, into estimates of food transport efficiency~\cite{lee_effects_2023}. Together, these results show that increasing curvature and roughness can reduce transport efficiency and increase congestion frequency, providing a mechanistic link from local geometry to system-level throughput.

\subsection{Traffic feedback on excavation and network remodeling}
In termite collectives, movement feeds back on substrate geometry through stigmergic coupling. This is the feedback loop sketched in Fig.~\ref{traffictermite}B: traffic creates local crowding and delay, these conditions can trigger construction, and the altered tunnel geometry changes future flow. At excavation fronts and along established routes, congestion can function as a signal for where remodeling is needed. Queue formation behind active excavators can trigger lateral digging, widening the tunnel and increasing capacity~\cite{barduniasQueueSizeDetermines2010}. Opposing excavation headings can likewise alter where deposition and excavation occur, influencing branch formation and the topology of the growing network~\cite{barduniasOpposingHeadingsExcavating2009}. Traffic thus does more than register resistance. Conduits that are narrow, contested, or slow can elicit remodeling. Only approximately 20\% of \textit{C.~formosanus} workers actively excavate at any given time~\cite{yangIndividualTaskLoad2009,barduniasBehavioralVariationTunnelers2010}, and agent-based models suggest that switching between digging and waiting states can generate coordination through neighbor-dependent transitions without requiring assessment of global colony state~\cite{otooleSelforganizedCriticalityTermite1999}. In fire ants, excavation in narrow tunnels is similarly optimized when a fraction of individuals remain idle, because idleness reduces flow-stopping clogs~\cite{aguilarCollectiveClogControl2018}. Under confinement, inactivity can act as a form of distributed inhibition that prevents saturation. Whether termite idleness plays a comparable role in regulating clogging and flow remains unresolved.

A second form of remodeling appears after resource discovery. Michael et al.\ showed that \textit{C.~formosanus} workers provided with pre-formed detour tunnels to food initially followed the existing route but later excavated shortcut branches that were then selectively widened~\cite{michaelFindingShortcutsCollective2023}. The colony thus restructures its transport network, shifting from an exploratory geometry suited to search to a geometry better suited to throughput. Movement determines which routes are reinforced, widened, or bypassed.

Taken together, these studies show that termite traffic interacts with soil structures. Encounters, delays, and bottlenecks are continually converted into changes in tunnel form and network use. Termites highlight a class of living transport systems in which transport and confinement coevolve: local flow conflicts actively reshape the geometry that governs future transport.

\section{Construction as substrate remodeling}

\begin{figure}
    \centering
    \includegraphics[width=1\linewidth]{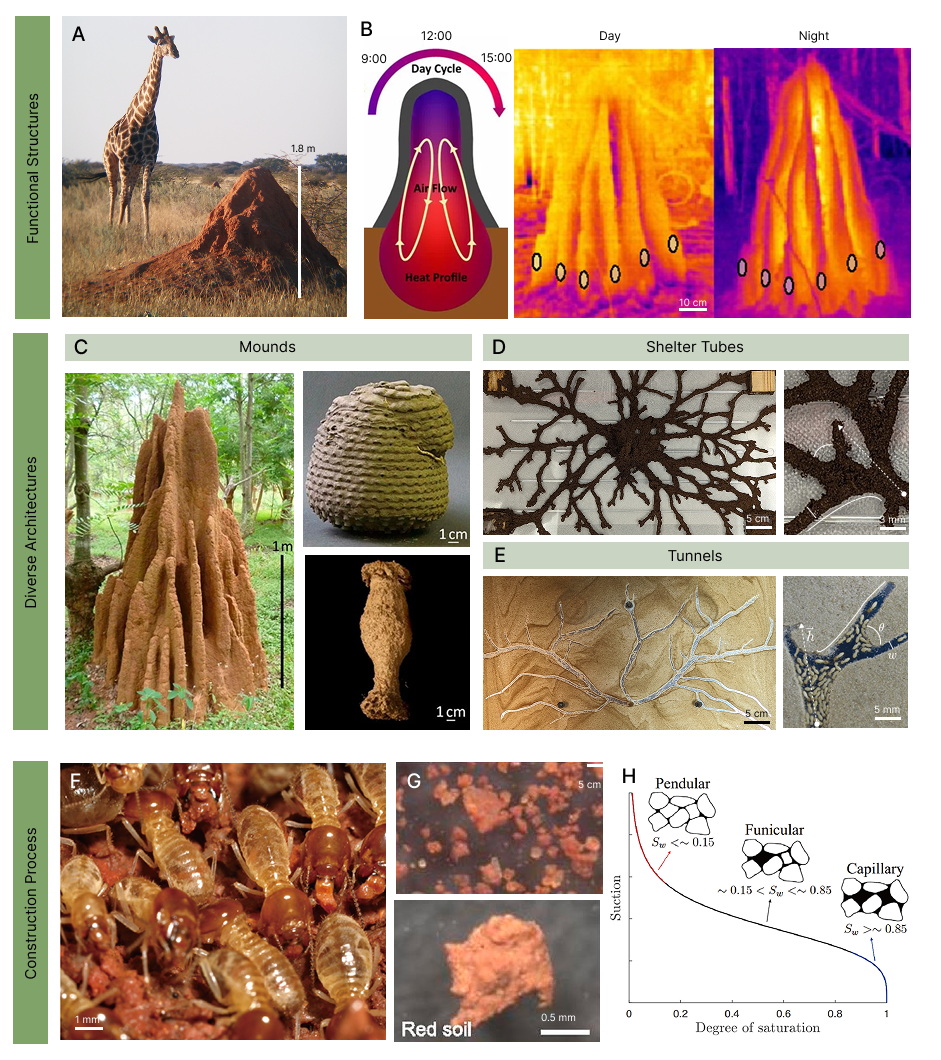} 
    \caption{\textbf{Termite construction processes and built structures across scales.} Termite-built structures range from large functional mounds to local transport conduits and material aggregates, all produced through repeated remodeling of granular substrates. (A) A \textit{Macrotermes natalensis} mound in Otjiwarongo, Namibia, illustrating the meter-scale architecture of fungus-growing termites (Photo: Paul Bardunias). (B) Thermal imaging and schematic representation of airflow and heat distribution in an \textit{Odontotermes obesus} mound, showing how mound geometry can mediate daily ventilation dynamics~\cite{kingTermiteMoundsHarness2015}. (C) Examples of mound diversity across taxa: an \textit{Odontotermes obesus} mound, an \textit{Apicotermes lamani} mound, and a \textit{Procornitermes araujoi} mound~\cite{ockoMorphogenesisTermiteMounds2019}. (D, E) Shelter-tube networks and excavated tunnel networks built by \textit{Coptotermes gestroi} in controlled laboratory settings (Photo: Kukhyun Lim). In both cases, local branch geometry can be quantified by variables such as branch width, branch length, inter-branch angle, and the homing vector of a branch tip. (F) \textit{Macrotermes} workers remodeling soil particles during construction. (G) Material boluses formed from red soil particles and termite secretion, illustrating the particle-scale units from which larger structures are assembled through a decentralized, additive construction process. (H) Soil-water retention curve showing pendular, funicular, and capillary regimes, which determine how liquid bridges and capillary forces mediate cohesion among particles during construction.}
    \label{fig:construction}
\end{figure}

A termite worker excavates or deposits soil pellets in response to cues available at the scale of its body, yet these local acts accumulate into structures that span multiple spatial scales. Across species (Fig.~\ref{fig:construction}A and~\ref{fig:construction}C), the visible mound is often only one part of a larger system that can include subterranean tunnels, above-ground shelter tubes, and mound conduits extending far beyond the central nest~\cite{King1969ForagingGO}. These structures regulate how the colony later moves, exchanges air, encounters moisture, and accesses food. The problem is both what termites build and how body-scale construction decisions generate architectures with colony-scale function.

\subsection{Structural modules and functional constraints}
Broad nesting categories such as one-piece, multiple-piece, and separate-piece nests remain useful for comparison across termite taxa~\cite{Abe1987EvolutionOL,mizumotoModernTermitesInherited2020,kikuchiNestingStrategyReflects2025}, but they do not by themselves capture how termite construction organizes local physical conditions. Across taxa, termite architecture can be understood more usefully as an assembly of functional modules. For the present review, the most relevant are tunnels and shelter tubes, which act as protected conduits for traffic and excavation; chambers, which serve as localized sites of exchange, storage, and interaction; and mound conduits, which form larger-scale channels for ventilation and thermal regulation~\cite{King1969ForagingGO,reinhardSystematicSearchFood1997,pernaStructureGalleryNetworks2008,kingTermiteMoundsHarness2015}. Together, these structures define a self-constructed hierarchy of spaces, from narrow confined passages to broader exchange and flow regions, in which geometry directly shapes sensing, transport, and subsequent remodeling.

Field measurements set the spatial scales over which these modules operate. 
Subterranean tunnel networks in \textit{C.~formosanus}, \textit{Reticulitermes flavipes}, and \textit{C.~travians} can span tens to more than a hundred meters laterally, reach meter-scale depths, and connect local centimeter-scale passages to foraging territories of hundreds to thousands of square meters~\cite{King1969ForagingGO,Su_Scheffrahn_1998,Lee2002ControlOF}. Mound-building species, by comparison, construct porous architectures at the meter scale, with heights of roughly $0.5$--$3$\,m and volumes on the order of $10$\,m$^3$ (Fig.~\ref{fig:construction}A and~\ref{fig:construction}C)~\cite{Miyagawa_Koyama_Kokubo_Matsushita_Adachi_Sivilay_Kawakubo_Oba_2011}. Mound diameter often covaries with height, suggesting approximately self-similar growth over much of the observed size range. These length and volume scales define the boundary conditions for physical models of termite construction: tunnels impose locally quasi-one-dimensional confinement within networks spanning tens of meters, mounds form three-dimensional porous structures at the meter scale, and surface constructions create thin interfaces between the colony and the external environment. How such remodeling and repair proceed depends on the local cues that guide excavation and deposition.

\subsection{Local cues that guide excavation and deposition}
Construction closes the loop of sensing and decision-making: workers read geometry and traffic, then imprint those signals back into the substrate. Excavation and deposition are guided by a small set of recurring cues, especially geometry, transport fields, crowding, and resource contact. These are local signals continually altered by the very activity they guide. Because they persist on different timescales, they couple slow structural change to faster transport and crowding dynamics.

\paragraph*{Geometry and transport fields.} Surface geometry provides the clearest example of a physical feature that is both sensed by workers and modified by their actions. In \textit{Macrotermes} mound construction, convex surfaces preferentially trigger material removal, whereas concave surfaces preferentially trigger deposition~\cite{caloviSurfaceCurvatureGuides2019}. A similar rule operates in \textit{C.~formosanus} tunnels, where protrusions are preferentially excavated and pellets are placed into concavities~\cite{lee_surface_2008} (Fig.~\ref{fig:construction}D and~\ref{fig:construction}E). Because each excavation or deposition event changes local curvature, the substrate records previous actions in a form that biases subsequent actions. Construction therefore becomes a curvature-sensitive feedback loop written directly into the growing structure.

Transport fields provide a second, less direct class of physical cues. Moisture, temperature, airflow, and gas gradients can influence both worker behavior and the material state of the substrate, but the links to local construction decisions are best resolved for moisture. In \textit{C.~formosanus}, high-moisture sand increases excavation volume~\cite{suTunnelingActivitySubterranean2003}, and workers transport moist soil from excavation sites onto adjacent dry surfaces, physically separating deposition from excavation~\cite{Carey_Calovi_Bardunias_Turner_Nagpal_Werfel_2019}. Because water applied during construction creates regions of elevated relative humidity that persist beyond the deposition event, moisture can act as a stigmergic template that marks active build zones, in addition to altering substrate mechanics and physiological conditions~\cite{Carey_Calovi_Bardunias_Turner_Nagpal_Werfel_2019}.

Airflow, temperature, and gas gradients are more clearly established at the scale of mound function. Mound-scale models suggest that advection and diffusion of heat and CO$_2$ through porous soil can couple strongly to architecture~\cite{ockoMorphogenesisTermiteMounds2019}, and field measurements show that diurnal temperature differences between inner and outer mound regions can drive convective ventilation through the conduit network (Fig.~\ref{fig:construction}B)~\cite{kingTermiteMoundsHarness2015}. Because excavation and deposition alter porosity, permeability, and pathway geometry, construction can in principle redirect these fluxes and thereby modify the fields that termites subsequently encounter. At present, however, curvature-sensitive excavation and deposition remain the clearest experimentally supported local construction rules in termites~\cite{caloviSurfaceCurvatureGuides2019,lee_surface_2008}. Moisture provides a second experimentally grounded substrate-mediated cue~\cite{suTunnelingActivitySubterranean2003,Carey_Calovi_Bardunias_Turner_Nagpal_Werfel_2019}, whereas the mapping from temperature, airflow, or gas fields onto specific local construction decisions remains less well resolved~\cite{ockoMorphogenesisTermiteMounds2019}.

\paragraph*{Crowding and resource contact.} Crowding provides a third signal (Fig.~\ref{fig:construction}E). Queue length at a tunnel tip directly affects tunnel width because waiting workers excavate laterally, widening the conduit or initiating a new branch~\cite{barduniasQueueSizeDetermines2010}. In de novo mound building by \textit{Macrotermes}, digging activity itself can recruit additional workers to crowded sites, creating aggregation without requiring a long-range attractant~\cite{greenExcavationAggregationOrganizing2017}. 

Resource distribution provides the fourth cue, though often only after contact has been made. In \textit{Reticulitermes flavipes}, above-ground wood elicits scouting, trail laying, and directed shelter-tube construction once discovered~\cite{reinhardSystematicSearchFood1997}. However, buried wood does not measurably alter tunneling direction before contact, even at millimeter-scale distances~\cite{pucheApplicationFractalAnalysis2001}. Resource location thus shapes construction primarily after discovery rather than guiding subterranean search at a distance.

\paragraph*{Material preparation and capillary binding.} The local rules described above operate within a material window set by the wet granular substrate. At the particle scale, termite construction is a wet-assembly process: workers collect soil particles, add secreted fluid, form moist boluses, and place them onto the growing structure (Fig.~\ref{fig:construction}F--\ref{fig:construction}H). Cohesion in these boluses arises largely from capillary bridges at particle contacts. In the pendular regime, isolated liquid menisci bind neighboring grains; as water content increases, bridges merge through funicular and capillary regimes, changing both cohesion and deformability~\cite{lian1993theoretical,mitarai2006wet}. A buildable bolus must therefore be wet enough to adhere, but not so wet that it slumps, a condition shaped by secretion volume, grain-size distribution, surface wettability, and ambient humidity~\cite{zachariah2017building}. Species differences in bolus composition, such as the mixing of salivary secretion with excavated granules in \textit{Macrotermes}, suggest that termites may tune this material window using taxon- and substrate-specific preparation strategies. Material preparation therefore does not provide a construction cue in the same sense as curvature or crowding; instead, it defines the physical conditions under which local excavation and deposition events persist as stable architecture.

These cues and material constraints are coupled through the soil itself. Excavation changes geometry, changed geometry redirects transport fields, redirected fields alter where workers accumulate, and crowding then biases the next construction event. Soil is the substrate through which information, moisture, gas exchange, and mechanical constraint propagate. In that sense, construction is the process by which the colony modifies the conditions under which later sensing and transport take place.

\subsection{From local rules to large-scale form}
If construction functions as control at the local scale, architecture is the accumulated record of that control at larger scales. Quantifying termite form makes it possible to ask which features of the final geometry preserve signatures of local interaction rules. In quasi-two-dimensional assays, \textit{C.~gestroi} shelter tubes branch more frequently than tunnels~\cite{kikuchiNestingStrategyReflects2025}, indicating that different structural modules can encode different construction dynamics even within the same species. The geometry of both modules can be described by a compact set of architectural variables, including branch width $w$, branch length $L$, inter-branch angle $\theta$, and a homing vector $\mathbf{h}$ associated with the direction of each growing tip (Fig.~\ref{fig:construction}D and~\ref{fig:construction}E). In laboratory assays, shelter tubes and tunnel networks grown by the same colony display broadly similar branching organization but differ systematically in these geometric descriptors~\cite{kikuchiNestingStrategyReflects2025}, providing a quantitative basis for distinguishing construction strategies even within a single species.

A similar logic applies to subterranean tunnels in \textit{C.~formosanus}. Branch angle and width are especially informative: branch angles cluster near values consistent with efficient branching flow~\cite{suCharacterizationTunnelingGeometry2004,barduniasOpposingHeadingsExcavating2009,Jackson2004TrailGG}, while tunnel width scales with queue length at the tip because waiting workers excavate laterally into the walls~\cite{barduniasQueueSizeDetermines2010}. In three dimensions, additional variables such as network tortuosity, connectivity, and conduit thickness may help reveal how mound and tunnel architectures balance wiring cost against transport benefit~\cite{ucarTheoryBranchingMorphogenesis2021a}. Large-scale form therefore appears to preserve signatures of local dynamics, although the mapping between specific interaction rules and architectural features remains incomplete.

Mechanistic models provide the bridge between local building rules and the later transport, exchange, and regulatory functions of the architecture. Box~2 summarizes a continuum view of this bridge, in which evolving geometry, transport fields, and material redistribution are treated as coupled coarse-grained variables. This approach is most developed at the mound scale, where porosity, thermal mass, and advective flux interact with construction to shape ventilation and morphology~\cite{ockoMorphogenesisTermiteMounds2019,kingTermiteMoundsHarness2015}. Field measurements show that diurnal thermal oscillations between inner and outer mound regions can drive convective airflow through the mound interior (Fig.~\ref{fig:construction}B)~\cite{kingTermiteMoundsHarness2015}. Complementary models that couple heat and gas transport to threshold-based construction rules reproduce the qualitative diversity of mound morphologies observed across species (Fig.~\ref{fig:construction}C)~\cite{ockoMorphogenesisTermiteMounds2019}. 

Discrete models address the complementary problem of how local rules accumulate into pillars, walls, or branching tunnel networks~\cite{bonabeauModelEmergencePillars1998,otooleSelforganizedCriticalityTermite1999,leeSimulationStudyTunnel2010}. Yet the bridge from local rules to empirically grounded mechanism remains incomplete. Much of the early modeling literature assumes diffusing chemical fields, even where empirical work now points more strongly to geometry-, crowding-, moisture-, and transport-mediated feedback. The central task is therefore to build models in which sensing, traffic, and construction are coupled through the same evolving substrate, with inputs that match the cues termites actually use and outputs that correspond to measurable architectural form.

\vspace{0.25in}
\centerline{\textbf{Box 2\ \textbar\ Continuum theories of construction: transport and morphogenesis}}
\par\noindent\rule{\textwidth}{0.4pt}
\begin{adjustwidth}{10pt}{10pt}
\small
    Continuum theories of construction seek to explain how repeated local acts of excavation and deposition generate large-scale form and function. Rather than tracking each worker individually, these approaches describe the evolving structure through coarse-grained fields such as interface position, material density, porosity, temperature, humidity, gas concentration, or flow. Their central question is how local rules for adding or removing material scale into persistent architectures that regulate transport, exchange, and environmental control~\cite{bonabeauModelEmergencePillars1998,ockoMorphogenesisTermiteMounds2019}. In the context of termite construction modules (Fig.~\ref{fig:construction}), these fields may describe tunnel walls, shelter-tube thickness, chamber boundaries, or mound conduit geometry across nested two- and three-dimensional spatial scales.

    The simplest description treats construction as motion of an interface. If \(h(\mathbf{x},t)\) denotes the local wall position, surface elevation, conduit radius, or thickness of a built structure, then
    \[
    \partial_t h = G(\mathbf{x},t)-L(\mathbf{x},t)
    \]
    where \(G\) and \(L\) are local growth and loss terms. Depending on the structure, \(h\) could represent the advancing boundary of a tunnel face, the thickening of a shelter tube, the reshaping of a chamber wall, or the outer surface of a mound. In termite collectives, the relevant rates may depend on worker activity, local crowding, curvature, moisture, or transport through the substrate~\cite{caloviSurfaceCurvatureGuides2019,ockoMorphogenesisTermiteMounds2019}.

    A second class of models makes this geometric dependence explicit:
    \[
    \partial_t h = F(\kappa,\nabla h,\rho,\phi,\ldots)
    \]
    where \(\kappa\) is curvature, \(\rho\) may denote local worker density, \(\phi\) a scalar field such as humidity or pheromone concentration, \(q\) a local traffic flux, and \(\ell\) a characteristic queue length or waiting-state measure. In this view, morphogenesis is a feedback process: shape alters local activity, and local activity rewrites shape.

    Construction can also be treated as a transport problem in a porous, evolving medium. If \(c(\mathbf{x},t)\) is a scalar field such as temperature, water vapor, or CO\(_2\) concentration, then
    \[
    \partial_t c + \nabla\cdot\mathbf{J}_c = S_c
    \qquad\text{and}\qquad
    \mathbf{J}_c = -D\nabla c + \mathbf{u}c
    \]
    where \(D\) is an effective diffusivity, \(\mathbf{u}\) an advective flow, and \(S_c\) a source or sink. In mound-scale theories, this framework has been used to understand how conduit geometry, porosity, and forcing shape ventilation, heat exchange, and gas transport~\cite{kingTermiteMoundsHarness2015,ockoMorphogenesisTermiteMounds2019}, where architecture regulates and contains the fields.

    These ideas suggest a local-to-global view of termite building. Local excavation and deposition alter geometry; altered geometry changes porosity, connectivity, and transport resistance; these changes redirect air, moisture, heat, and traffic; and the resulting fields feed back on where additional work is likely to occur. A useful continuum theory would bridge at least three levels: local building rules, intermediate geometric variables such as curvature, width, porosity, or connectivity, and global functional consequences such as ventilation, thermoregulation, or transport efficiency~\cite{ockoMorphogenesisTermiteMounds2019}.

    There is predictive power to continuum theories for termite studies. For example, sharp corners should smooth over time under curvature-biased excavation and deposition. Widening rates at tunnel bottlenecks or excavation fronts may increase once crowding or queue length exceeds a threshold~\cite{barduniasQueueSizeDetermines2010}. Moisture should shift excavation-front speed, branch persistence, and local deposition patterns~\cite{suTunnelingActivitySubterranean2003}. More generally, termite construction highlights a broader problem in active matter and biological morphogenesis: how to describe agents that continuously tune the transport and regulatory properties of the self-constructed medium they inhabit.
\end{adjustwidth}
\par\noindent\rule{\textwidth}{0.4pt}

\section{Synthesis and outlook}
Two limitations frame the current literature. First, most mechanistic evidence comes from a small number of subterranean termites, especially \textit{Coptotermes spp.}, studied in planar or otherwise simplified tunnel assays. Second, the bridge from worker-scale rules to mound- or network-scale architecture remains much stronger in theory than in direct measurement.

Both limitations are now becoming more tractable. Transparent soil systems, X-ray microtomography, high-speed videography, and machine-learning-based tracking and behavioral classification increasingly allow simultaneous observation of individual workers and the substrate they reshape in three-dimensional, deformable, and more naturalistic environments~\cite{karibi-botoyeTermiteMoundArchitecture2025,downieTransparentSoilImaging2012,sharmaTransparentSoilMicrocosms2020,minterMorphogenesisExtendedPhenotype2012,scloccoIntegratingRealtimeData2021}. At the same time, inference tools from statistical physics and information theory, including transfer entropy and related approaches for reconstructing interaction structure from noisy trajectory, contact, and transport data, provide new ways to connect movement, encounters, and substrate modification to the local rules that generate them~\cite{poissonnierExperimentalInvestigationAnt2019,basakTransferEntropyDependent2021,chasePhysicsSensingDecisionMaking2025a}.

These developments make several questions newly tractable. What geometric features of tunnel surfaces are encoded by the termite antenna, and over what spatial and temporal scales? What signals trigger transitions among exploratory excavation, transport, widening, branching, merging, reinforcement, and repair? How do congestion, worker density, traffic asymmetries, and substrate geometry interact to determine whether interference is tolerated, behaviorally resolved, or converted into remodeling? Which architectural variables best preserve the signatures of local rules across species, substrates, and stages of construction? And can the diversity of termite nests be organized by a small set of physical and behavioral control parameters, or does it reflect fundamentally different construction strategies?

These same developments also sharpen the modeling task. Much of the older literature treats diffusing chemical fields as the primary coupling mechanism, even where empirical work now points more strongly to geometry-, crowding-, and transport-mediated feedback, and in some cases to direct mechanical interactions with the substrate itself. The next step is to build models whose inputs match the signals termites actually encounter and alter, and whose predictions can be tested against trajectories, contact patterns, transport statistics, and evolving architecture.

Taken together, the literature supports three broad conclusions. First, termite sensing is predominantly contact-based and scaffolded by geometry, but workers can also integrate their own movements to maintain direction beyond the immediately local scale. Second, termite traffic is a feedback process in which congestion, delay, and rerouting can reshape the network itself. Third, construction functions as substrate remodeling, because excavation and deposition continuously rewrite the geometry through which later information, movement, and coordination propagate.

Across sensing, movement, and construction, termites act locally while continuously rewriting the substrate through which later information and transport propagate. The central challenge is to connect worker-scale interactions to colony-scale architecture and function. More broadly, termites exemplify a wider class of living active systems in which agents construct the geometry, transport pathways, and information-bearing substrate that shape their own future dynamics. The theoretical task is to extend active-matter descriptions beyond motion in prescribed confinement toward frameworks in which confinement, transport resistance, and environmental memory are themselves emergent dynamical variables. Subterranean and mound-building termites thus provide a tractable system for studying how collective behavior, material transport, and substrate modification become coupled across scales to generate robust architecture without centralized control~\cite{prasathDynamicsCooperativeExcavation2022,werfelDesigningCollectiveBehavior2014}.

\bibliographystyle{unsrt}

\end{document}